\documentclass[prd,nofootinbib]{revtex4}
\usepackage{amsmath}
\usepackage{amssymb}
\usepackage{epsf}

\begin{document}

\title{Gravitational radiation reaction in compact binary systems: \\
Contribution of the magnetic dipole-magnetic dipole interaction}
\author{M\'{a}ty\'{a}s Vas\'{u}th$^{1}$, Zolt\'{a}n Keresztes$^{2}$, Andr%
\'{a}s Mih\'{a}ly$^{2}$, and L\'{a}szl\'{o} \'{A}. Gergely$^{2,3}$}
\affiliation{}
\affiliation{$^{1}$KFKI Research Institute for Particle and Nuclear Physics, Budapest
114, P.O.Box 49, H-1525 Hungary}
\affiliation{$^2$Astronomical Observatory and Department of Experimental Physics,
University of Szeged, Szeged 6720, D\'{o}m t\'{e}r 9, Hungary}
\affiliation{$^3$Institute of Cosmology and Gravitation, University of Portsmouth,
Portsmouth PO1 2EG, UK }

\begin{abstract}
We study the gravitational radiation reaction in compact binary systems
composed of neutron stars with spin and huge magnetic dipole moments
(magnetars). The magnetic dipole moments undergo a precessional motion about
the respective spins. At sufficiently high values of the magnetic dipole
moments, their interaction generates second post-Newtonian order
contributions both to the equations of motion and to the gravitational
radiation escaping the system. We parametrize the radial motion and average
over a radial period in order to find the secular contributions to the
energy and magnitude of the orbital angular momentum losses, in the generic
case of \textit{eccentric} orbits. Similarly as for the spin-orbit,
spin-spin, quadrupole-monopole interactions, here too we deduce the secular
evolution of the relative orientations of the orbital angular momentum and
spins. These equations, supplemented by the evolution equations for the
angles characterizing the orientation of the dipole moments form a first
order differential system, which is closed. The circular orbit limit of the
energy loss agrees with Ioka and Taniguchi's earlier result.
\end{abstract}

\date{\today}
\startpage{1}
\maketitle

\section{Introduction}

Neutron star and black hole binary systems are among the most probable
sources of the gravitational radiation emitted in the frequency range of the
Earth-based interferometric detectors such as the Laser Interferometric
Gravitational Wave Observatory (LIGO) \cite{LIGO}, VIRGO \cite{VIRGO}, GEO 
\cite{GEO}, and TAMA \cite{TAMA} (all will detect in the High-Frequency Band:%
$\ 1$ Hz to $10^{4}$ Hz) and the envisaged Laser Interferometer Space
Antenna (LISA)\cite{LISA,LISA1}.(Low-Frequency Band: $10^{-4}$ Hz to$\ 1$
Hz) respectively. A post-Newtonian (PN) treatment of $3.5$ PN orders is
generally agreed to describe with sufficient accuracy both the motion and
the gravitational radiation up to the point of the innermost circular orbit
(ICO), at least in the case of neutron star binaries \cite{Blanchet}. While the
3PN order approximation is expected to locate the ICO with an accuracy of
1\% for binary systems with comparable masses, it breaks down in the
region inner to the ICO in the latest stage of the final coalescence, when
numerical relativity is required. However, there is a gap between the
failure of the PN expansion and the beginning of the merger, this being
called the intermediate binary black hole (IBBH) problem \cite{IBBH}, recently 
assessed in \cite{Buonanno1}. 

The effort of providing templates for the search of the gravitational waves
in the noisy background is underway, at least for binaries on circular
orbits, characterized by their mass and the radius of the orbit. The
assumption of circular orbits is justified by the circularizing property of
the gravitational waves \cite{Peters}, still interesting situations can
occur in galactic nuclei, when freshly formed binary systems have not enough
time to circularize \cite{QuSha,HiBe}. The error produced by the neglection
of the eccentricity during the search for gravitational waves was estimated
to be substantial \cite{PoMa}. The assumption of disregarding other
characteristics of the binary components, like the spins, quadrupole moments
and even magnetic dipole moments (to be discussed in the present paper), in
many cases is not well founded. For example there is increasing
observational evidence for an overhelming majority of galactic black holes
(even colliding ones) with maximally allowed spins \cite{Biermann}. While
neutron star binaries are not expected to be significantly spinning \cite%
{nospin1}, \cite{nospin2}, neutron stars with super-strong magnetic fields
(magnetars) have also been detected \cite{magnetars}.

Therefore a generic treatment, allowing eccentric orbits and all types of
contributions up to the $3.5^{th}$PN order would be desirable in the
PN-description of gravitational waves. The complexity of such results
forbids their immediate implementation in templates, however with increasing
computational capacity\ this is hoped to become possible. An other
complicating feature arising in the eccentric case is the occurence of
several harmonics in the incoming gravitational wave which, contrarely to
the circular case (where the dominant frequency is twice of the orbital
frequency) will have comparable intensities \cite{PetersMathews}. In the
search for precessing binaries the use of effective templates has been
proposed recently \cite{Buonanno2}, \cite{Grandclement}.

The eccentricity was taken into account in the second PN order description
of \cite{GI}. The 1.5 PN spin-orbit and 2PN spin-spin contributions to the
orbital evolution and gravitational radiation reaction were considered in 
\cite{BOC,ACST,KWW,Kidder,RS,GPV3,spinspin1,spinspin2}. The gravitational
quadrupole-monopole type radiative contribution was also treated recently 
\cite{quadrup}, the results matching the previous circular orbit computation 
\cite{Poisson}.

The effect of possible strong magnetic fields of the compact binary
components was discussed by Ioka and Taniguchi \cite{IT}, for the case of
circular orbits. They argue that in the upper limit of $10^{16}$ $G$ for the
magnetic fields the coupling of the magnetic dipole moments generates
contributions of the same magnitude than the 2PN corrections to both the
equation of motion and the gravitational radiation. They also compute the
corresponding electromagnetic radiation, and find out that the
electromagnetically radiated power is \textquotedblright much
less\textquotedblright\ than the power radiated away gravitationally. The
ratio of the two types of radiations is $3/64$ cf. Eq. (16) of \cite{IT},
which counts approximately half of a PN order. Being interested in the 
\textit{leading order radiation due to the magnetic dipoles,} we will
disregard the electromagnetic component of radiation. Whether the leading
order gravitational radiation due to magnetic moments truely appears at
second or only at higher orders depends on the strength of the magnetic
fields. Actual observational evidence supports magnetic fields of $10^{15}$ $%
G$ for isolated magnetars \cite{magnetars}.

In the present paper we would like to extend our previous computations
carried on for the spin-orbit (SO), spin-spin (SS) and quadrupole-monopole
(QM) interactions for the case of the magnetic dipole - magnetic dipole
interaction. This would mean a generalization of Ioka and Taniguchi's
description allowing for eccentric orbits.

In Sect. II we set up the formalism, by introducing the generalized true and
eccentric anomaly parametrizations for the magnetic dipole - magnetic dipole
perturbation. As proved earlier \cite{param}, all relevant integral
expressions can be easily computed by use of the residue theorem, with the
additional bonus that in the majority of cases the only pole is at the
origin. Our treatment follows closely Refs. \cite{spinspin1,spinspin2} and 
\cite{quadrup}. In contrast with the energy $E$, the magnitude of the
orbital angular momentum is not a conserved quantity in the absence of
radiation at this order. Thus we introduce its angular average $\bar{L}$ in
order to characterize the perturbed radial motion. The radial period, as
well as the relation between $\bar{L}$ and the time average $\langle
L\rangle $ resemble the results of the previously discussed cases.

Sect. III \ contains the main results of the paper. These are the secular
evolutions of $E$ and $\bar{L}$ due to the magnetic dipole - magnetic dipole
contribution\ to the gravitational radiation \ Also, as in previous cases,
the evolution of the angle variables $\kappa _{i}$ $\ $and $\gamma $, which
characterize the relative orientation of the spin vectors and orbital
angular momentum vector are derived. However the evolution of this set of
variables $\left( E\text{, }\bar{L},\text{ }\kappa _{i},\text{ }\gamma
\right) $ does not close to a first order differential system, as in
previous cases. In order to close the system, we have to compute the
evolution equations for the angle variables $\beta _{i}$ characterizing the
relative orientation of the spins and magnetic dipole moments. The
reliability of our results is checked in the Concluding Remarks, where in
the circular limit we recover the expression for the energy loss given in 
\cite{IT}.

The velocity of light $c$ and the gravitational constant $G$ are kept in all
expressions.

\section{The radial motion}

We consider a binary system composed of neutron stars with magnetic dipole
moments $\mathbf{{d}_{i}}$ (properly scaled in order to absorb all
dimension-carrying constants). The dipole-dipole interaction is given by the
Lagrangian \cite{IT}: 
\begin{equation}
\mathcal{L}=\mathcal{L}_{N}+\mathcal{L}_{DD}=\frac{\mu }{2}\mathbf{v}^{2}+%
\frac{Gm\mu }{r}+\frac{1}{r^{3}}\left[ 3(\mathbf{n\cdot d_{1}})(\mathbf{%
n\cdot {d}_{2}})-\mathbf{d_{1}\cdot {d}_{2}}\right] \ ,
\end{equation}
where $\mathbf{v}=\mathbf{\dot{r}}$ and $\mathbf{r}=r\mathbf{n}$.
Accordingly, the total acceleration is the sum of the Newtonian $\mathbf{a}%
_{N}$ and the dipole-dipole contribution $\mathbf{a}_{DD}$: 
\begin{eqnarray}
\mathbf{a} &=&\mathbf{a}_{N}+\mathbf{a}_{DD}\ ,  \notag \\
\mathbf{a}_{N} &=&-\frac{Gm}{r^{2}}\mathbf{n}\ ,  \notag \\
\mathbf{a}_{DD} &=&\frac{3}{\mu r^{4}}\left\{ \left[ \mathbf{{d}_{1}\cdot {d}%
_{2}}-5(\mathbf{n\cdot {d}_{1}})(\mathbf{n\cdot {d}_{2}})\right] \mathbf{n}%
+\sum_{i\neq j}(\mathbf{n\cdot {d}_{i}})\mathbf{d}_{\mathbf{j}}\right\} \ .
\end{eqnarray}
The orbital angular momentum $\mathbf{L}=\mathbf{r\times p}=\mathbf{L}_{N}$
\ evolves due to the dipole-dipole perturbation as 
\begin{equation}
\mathbf{\dot{L}}=\frac{3}{r^{3}}\sum_{i\neq j}(\mathbf{n\cdot {d}_{i}})(%
\mathbf{n\times {d}_{j}})\ ,
\end{equation}
and its magnitude $L$ changes accordingly 
\begin{equation}
\dot{L}=\mathbf{\dot{L}}\mathbf{\hat{L}}=\frac{3\mu }{Lr^{2}}\sum_{i\neq j}(%
\mathbf{n\cdot {d}_{i}})(\mathbf{v}-\dot{r}\mathbf{n)\cdot {d}_{j}}\ .
\label{lmagdot}
\end{equation}
The energy $E=\dot{\mathbf{r}}\partial \mathcal{L}/\partial \dot{\mathbf{r}}-%
\mathcal{L}$ and the total angular momentum $\mathbf{J=L+S_{1}+S_{2}}$ are
the constants of this motion.

\begin{figure}[tbh]
\epsfysize=9cm \centerline{\hfill
\epsfbox{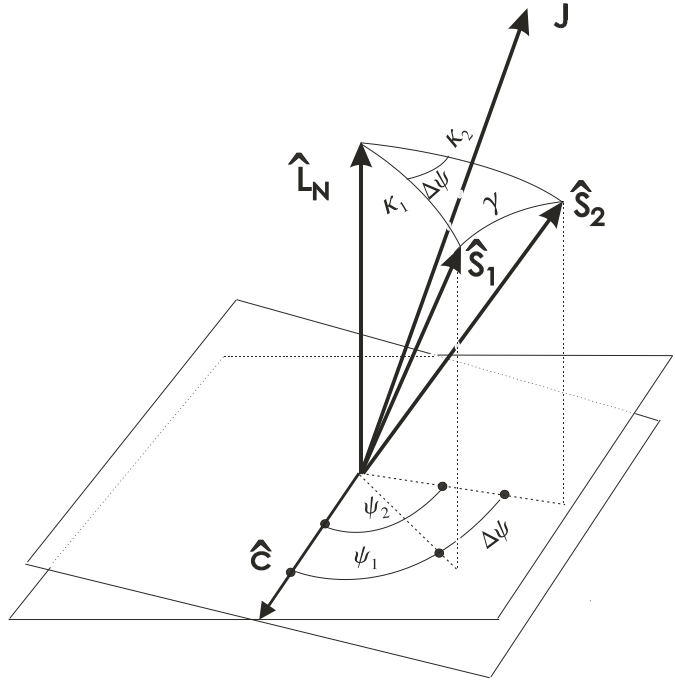}\hfill}
\caption{The relative angles $\protect\kappa_i$ and $\protect\gamma$ of the $%
\mathbf{L_{N}}$ and $\mathbf{S_{i}}$ angular momenta and the azimuthal
angles $\protect\psi_{i}$ of the spins measured in the plane perpendicular
to $\mathbf{L}$. The vector $\mathbf{\hat{c}}$ lies on the intersection of
this plane with the one perpendicular to $\mathbf{J}$.}
\end{figure}

We introduce three orthonormal coordinate systems $\mathcal{K}$ and $%
\mathcal{K}^{i}$ with the axes $(\mathbf{\hat{c},\hat{L}\times \hat{c},\hat{L%
}})$ and $(\mathbf{\hat{b}_{i},\hat{S}_{i}\times \hat{b}_{i},\hat{S}_{i}})$,
where $\mathbf{\hat{c}}$ and $\mathbf{\hat{b}_{i}}$ are the unit vectors in
the $\mathbf{J\times L}$ and $\mathbf{S_{i}\times L}$ directions,
respectively. In the system $\mathcal{K}$ the polar angles $\kappa _{i}$ and 
$\psi _{i}$ of the spins are defined as $\mathbf{\hat{S}_{i}=}(\sin \kappa
_{i}\cos \psi _{i},\sin \kappa _{i}\sin \psi _{i},\cos \kappa _{i})$ (Fig
1). We also introduce the polar angles $\alpha _{i}$ and $\beta _{i}$ of the
the magnetic dipole moments $\mathbf{{d}_{i}\,}$in the corresponding system $%
\mathcal{K}^{i}$ as $\mathbf{{\hat{d}}_{i}=}(\sin \alpha _{i}\cos \beta
_{i},\sin \alpha _{i}\sin \beta _{i},\cos \alpha _{i})$ (Fig 2). The
transformation $\mathcal{K}^{i}\rightarrow \mathcal{K}$ represents a
sequence of rotations $R_{z}(-\tau _{i})R_{x}(-\kappa _{i})$, where the
angles $\tau _{i}=\cos ^{-1}(\mathbf{\hat{c}\cdot \hat{b}_{i}})$ satisfy the
relations $\tau _{i}+\psi _{i}=\pi /2$ (see Fig. 2 for a proof). Thus in the 
$\mathcal{K}$ system the vectors appearing in Eq. (\ref{lmagdot}) become%
\footnote{%
Due to the magnetic dipole - magnetic dipole perturbation all three of the
above defined coordinate systems may evolve (to $DD$-order). We do not
consider these evolutions, as the components of the above vectors enter
solely in various $DD$-terms, thus they are needed with Newtonian accuracy.}%
: 
\begin{eqnarray}
\mathbf{n}=\left( 
\begin{array}{c}
\cos\psi \\ 
\sin\psi \\ 
0%
\end{array}
\right) \ &,&\quad \mathbf{v}=\dot{r}\mathbf{n}+\frac{L}{\mu r} \left( 
\begin{array}{c}
-\sin\psi \\ 
\cos\psi \\ 
0%
\end{array}%
\right) \ ,\quad  \notag \\
\mathbf{{d}_{i}} &=&d_{i}\left( 
\begin{array}{c}
\rho_i\sin\psi_i + \sigma_i\cos\psi_i \\ 
\sigma_i\sin\psi_i - \rho_i\cos\psi_i \\ 
\zeta_i%
\end{array}%
\right) \ ,  \label{coords}
\end{eqnarray}
where we have introduced the shorthand notations 
\begin{eqnarray}
\rho _{i} &=&\sin \alpha _{i}\cos \beta _{i}\ ,  \notag \\
\nu _{i} &=&\sin \alpha _{i}\sin \beta _{i}\ ,  \notag \\
\sigma _{i} &=&\cos \alpha _{i}\sin \kappa _{i}+\nu _{i}\cos \kappa _{i}\ , 
\notag \\
\zeta _{i} &=&\cos \alpha _{i}\cos \kappa _{i}-\nu _{i}\sin \kappa _{i}\ .
\end{eqnarray}

\begin{figure}[tbh]
\epsfysize=9cm \centerline{\hfill
\epsfbox{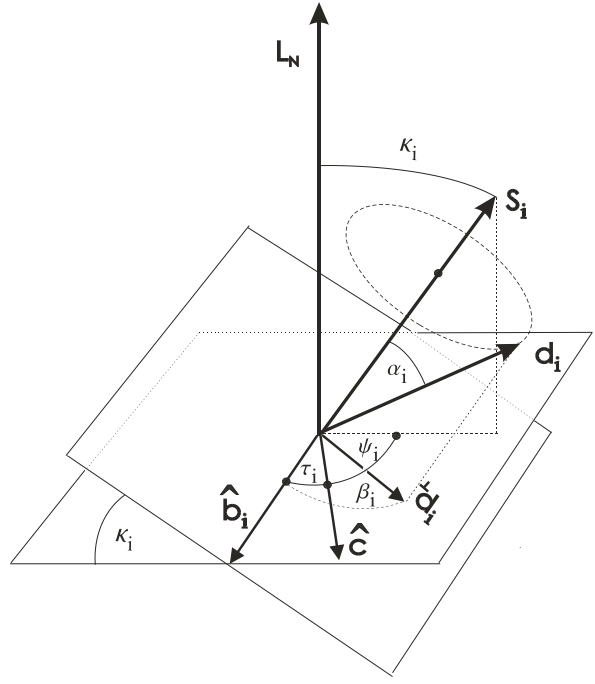}\hfill}
\caption{The magnetic dipole moment of each neutron star undergoes a
precessional motion about the spin vector of the respective neutron star.
The projection of the magnetic dipole moment into the plane perpendicular to 
$\mathbf{S_{i}}$ is denoted $^{\perp }\mathbf{d}_{\mathbf{i}}$, a vector
characterized by the azimuthal angle ${\protect\beta_i}$. On the
intersection of the planes perpendicular to $\mathbf{L_{N}}$ and $\mathbf{%
S_{i}}$ lies the vector $\mathbf{{\hat b}_{i}}$. As the projection of $%
\mathbf{S_{i}}$ into the plane of motion lies in the plane defined by $%
\mathbf{S_{i}}$ and $\mathbf{L_{N}}$, a plane to which the vector $\mathbf{%
\hat{b_i}}$ is perpendicular by definition, the relation ${\protect\tau_i+%
\protect\psi_i=\protect\pi /2}$ holds.}
\end{figure}

Since $\mathcal{L}_{DD}$ is independent of the velocity, $E_{DD}=-\mathcal{L}%
_{DD}$ holds: 
\begin{equation}
E_{DD}=\frac{d_{1}d_{2}}{2r^{3}}\left[ \mathcal{A}_{0}-3\mathcal{B}_{2}(\chi
)\right] \ .  \label{Edd}
\end{equation}
Here we have introduced the Newtonian true anomaly parameter $\chi =\psi
-\psi _{0}$ and 
\begin{eqnarray}
\mathcal{A}_{0} &=&2\cos \lambda +3(\rho _{1}\sigma _{2}-\rho _{2}\sigma
_{1})\sin (\delta _{1}-\delta _{2})-3(\rho _{1}\rho _{2}+\sigma _{1}\sigma
_{2})\cos (\delta _{1}-\delta _{2})\ ,  \label{A0} \\
\mathcal{B}_{k}(\chi ) &=&(\sigma _{1}\sigma _{2}-\rho _{1}\rho _{2})\cos
(k\chi +\delta _{1}+\delta _{2})-(\rho _{1}\sigma _{2}+\rho _{2}\sigma
_{1})\sin (k\chi +\delta _{1}+\delta _{2})\ ,  \label{Bk}
\end{eqnarray}
where $\lambda $ is the angle subtended by the magnetic moments and $\delta
_{i}=\psi _{0}-\psi _{i}$. We stress here that he angular average of (\ref%
{Bk}) vanishes $\left\langle \mathcal{B}_{k}(\chi )\right\rangle =0$ for any
integer $k$.

With these notations, the evolution of the magnitude of angular orbital
momentum is: 
\begin{equation}
\dot{L}=\frac{3d_{1}d_{2}}{2r^{3}}\mathcal{B}_{2}^{\prime }(\chi )\ ,
\end{equation}
where a prime denotes derivative with respect to $\chi $. In all DD
expressions $\rho _{i}$ and $\sigma _{i}$ can be regarded as constants,
since the evolution of the angles $\beta _{i}$ and $\kappa _{i}$ are of
1.5PN order and $\alpha _{i}$ are constants. Similar considerations hold for 
$\lambda $. After the integration $\int \dot{L}\dot{\chi}^{-1}d\chi $, we
obtain 
\begin{eqnarray}
L(\chi ) &=&L_{0}-\frac{\mu ^{2}d_{1}d_{2}}{2\bar{L}^{3}}\bigl\{(3Gm\mu +4%
\bar{A})\mathcal{B}_{0}  \notag \\
&&-(3Gm\mu +4\bar{A}\cos \chi )\mathcal{B}_{2}(\chi )+\bar{A}\sin \chi 
\mathcal{B}_{2}^{\prime }(\chi )\bigr\}\ ,  \label{Lchieq}
\end{eqnarray}
%
where $\bar{A}$ is the magnitude of the Laplace-Runge-Lenz vector for a
Keplerian motion characterized by $E$ and $\bar{L}$ 
\begin{equation}
\bar{A}=\left( G^{2}m^{2}\mu ^{2}+\frac{2E\bar{L}^{2}}{\mu }\right) ^{1/2}\ .
\end{equation}
From (\ref{Lchieq}) we see that $L(0)=L(2\pi )=L_{0}$ holds. Let $\bar{L}$
denote the angular average of $L(\chi )$ 
\begin{equation}
\bar{L}=L_{0}-\frac{\mu ^{2}d_{1}d_{2}}{2\bar{L}^{3}}(3Gm\mu +4\bar{A})%
\mathcal{B}_{0}\ .  \label{Lbareq}
\end{equation}
Then we can express $L(\chi )$ in terms of $\bar{L}$ rather than $L_{0}:$ 
\begin{eqnarray}
L(\chi ) &=&\bar{L}+\delta L_{DD}\ ,  \label{Lchi} \\
\delta L_{DD} &=&\frac{\mu ^{2}d_{1}d_{2}}{2\bar{L}^{3}}\left[ (3Gm\mu +4%
\bar{A}\cos \chi )\mathcal{B}_{2}(\chi )-\bar{A}\sin \chi \mathcal{B}%
_{2}^{\prime }(\chi )\right] \ .  \label{deltaL}
\end{eqnarray}

The magnitude of the velocity and the radial equation take the form 
\begin{eqnarray}
v^{2} &=&\frac{2}{\mu }\left[ E-E_{DD}(r,\chi )\right] +\frac{2Gm}{r}\ ,
\label{veloc} \\
\dot{r}^{2} &=&\frac{2}{\mu }\left[ E-E_{DD}(r,\chi )\right] +\frac{2Gm}{r}-%
\frac{L(\chi )^{2}}{\mu ^{2}r^{2}}\ .  \label{radeq}
\end{eqnarray}
Turning points are defined as solutions of the $\dot{r}=0$ equation. They
can be expressed as 
\begin{equation}  \label{rpmgen}
r_{{}_{{}_{\min }^{\max }}}=r_{\pm }\mp \frac{\mu r_{\pm }^{2}E_{DD}^{\pm }+%
\bar{L}\delta L_{DD}^{\pm }}{\mu \bar{A}}\ ,
\end{equation}
where $E_{DD}^{\pm }=E_{DD}\left( r_{\pm }\right) $, $\delta L_{DD}^{\pm
}=\delta L_{DD}\left( r_{\pm }\right) $ and $r_{\pm }$ are the turning
points for a Keplerian orbit characterized by $E$ and $\bar{A}$.
Substituting (\ref{Edd}) and (\ref{deltaL}) we obtain 
\begin{equation}
r_{{}_{{}_{\min }^{\max }}}=\frac{Gm\mu \pm \bar{A}}{-2E}+\frac{\mu
d_{1}d_{2}}{2\bar{A}\bar{L}^{2}}\left\{ (\bar{A}\mp Gm\mu )\mathcal{A}_{0}+%
\bar{A}\mathcal{B}_{0}\right\} \ .  \label{turn}
\end{equation}
Next we introduce the generalized true and eccentric anomaly
parametrizations of the radial motion, following the generic receipe from
Ref. \cite{param}: 
\begin{eqnarray}
r &=&\frac{\bar{L}^{2}}{\mu (Gm\mu +\bar{A}\cos \chi )}+\frac{\mu
d_{1}d_{2}\Lambda }{2\bar{A}\bar{L}^{2}(Gm\mu +\bar{A}\cos \chi )^{2}}\ , \\
\Lambda &=&\bar{A}\left[ (3G^{2}m^{2}\mu ^{2}+\bar{A}^{2})\mathcal{A}%
_{0}+(G^{2}m^{2}\mu ^{2}+\bar{A}^{2})\mathcal{B}_{0}\right]  \notag \\
&&+Gm\mu \left[ (G^{2}m^{2}\mu ^{2}+3\bar{A}^{2})\mathcal{A}_{0}+2\bar{A}^{2}%
\mathcal{B}_{0}\right] \cos \chi \ , \\
r &=&\frac{Gm\mu -\bar{A}\cos \xi }{-2E}+\frac{\mu d_{1}d_{2}}{2\bar{A}\bar{L%
}^{2}}\left\{ \bar{A}\left( \mathcal{A}_{0}+\mathcal{B}_{0}\right) +Gm\mu 
\mathcal{A}_{0}\cos \xi \right\} \ .
\end{eqnarray}
Introducing suitable complex variables, radial expressions can be averaged
by use of the residue theorem, as described in \cite{param}. Following this
recipe, the radial period turns out to have the Keplerian form 
\begin{equation}
T=2\pi Gm\left( \frac{\mu }{-2E}\right) ^{3/2}\ .
\end{equation}
(This holds whenever the radial dependence of the perturbing terms in the
radial equation (\ref{radeq}) is either $1/r^{2}$ or $1/r^{3}$ \cite{param}%
.) Time average of $L(\chi )$ gives 
\begin{equation}
\langle L\rangle ={\bar{L}}+\frac{d_{1}d_{2}\mathcal{B}_{0}}{2\mu \bar{A}^{2}%
\bar{L}^{3}}\left[ Gm\mu ^{4}(2G^{2}m^{2}\mu ^{2}-3{\bar{A}}^{2})-2\left(
-2\mu E\right) ^{3/2}{\bar{L}}^{3}\right] \ .  \label{LaveLbar}
\end{equation}
This latter expression allows for an equivalent expression of all
forthcoming results in terms of $\langle L\rangle $ rather than ${\bar{L}}$,
case needed. We remark that the square bracket in Eq. (\ref{LaveLbar})
coincides with the factor $F_{1}/F_{2}$ of the corresponding expressions in
the quadrupole-monopole and spin-spin cases, Refs. \cite{quadrup} and \cite%
{spinspin1}.

\section{Leading order magnetic dipole-magnetic dipole contribution to the
evolutions of the dynamical variables under radiation reaction}

\subsection{Energy loss}

The radiative change in the energy to leading order is given by the
quadrupole formula 
\begin{equation}
\frac{dE}{dt}=-\frac{G}{5c^{5}}I^{(3)jl}I^{(3)jl}\ ,
\end{equation}
where $I^{(3)jl}$ is the $3^{rd}$ time derivative of the system's symmetric
trace-free (STF) mass quadrupole moment tensor. To leading order it is given
as 
\begin{equation}
I_{N}^{jl}=\mu \left( x^{j}x^{l}\right) ^{STF}\ .
\end{equation}
Inserting $v^{2}$ and $\dot{r}^{2}$ from (\ref{veloc}) and (\ref{radeq}),
together with $E_{DD}$ and $L(\chi )$ from (\ref{Edd}) and (\ref{Lchi}) into
the quadrupole formula we obtain 
\begin{eqnarray}
\frac{dE}{dt} &=&\left( \frac{dE}{dt}\right) _{N}+\left( \frac{dE}{dt}%
\right) _{DD}\ , \\
\left( \frac{dE}{dt}\right) _{N} &=&-\frac{8G^{3}m^{2}}{15c^{5}r^{6}}\left(
2E\mu r^{2}+2Gm\mu ^{2}r+11\bar{L}^{2}\right) \ , \\
\left( \frac{dE}{dt}\right) _{DD} &=&\frac{4G^{2}md_{1}d_{2}}{15c^{5}\mu 
\bar{L}^{2}r^{8}}\left\{ \sum_{k=1}^{3}a_{k}\mathcal{B}_{k}(\chi )+a_{4}%
\mathcal{A}_{0}\right\}
\end{eqnarray}
with the coefficients $a_{k}$ given by 
\begin{eqnarray}
a_{1} &=&3\mu \bar{A}r(-22Gm\mu ^{2}r+17\bar{L}^{2})\ ,  \notag \\
a_{2} &=&6(-11G^{2}m^{2}\mu ^{4}r^{2}+6E\bar{L}^{2}\mu r^{2}+5Gm\mu ^{2}\bar{%
L}^{2}r-51\bar{L}^{4})\ ,  \notag \\
a_{3} &=&-\mu \bar{A}r(22Gm\mu ^{2}r+51\bar{L}^{2})\ ,  \notag \\
a_{4} &=&2\bar{L}^{2}(-6E\mu r^{2}-5Gm\mu ^{2}r+39\bar{L}^{2})\ .
\end{eqnarray}
To compute the averaged loss of the energy we parametrize the above
expression by the true anomaly $\chi $ and pass to the complex variable $z=%
\mathrm{exp}(i\chi )$. The only pole of the integrand is located in the
origin. Using the residue theorem we obtain 
\begin{eqnarray}
\left\langle \frac{dE}{dt}\right\rangle &=&\left\langle \frac{dE}{dt}%
\right\rangle _{N}+\left\langle \frac{dE}{dt}\right\rangle _{DD}\ ,
\label{averEloss} \\
\left\langle \frac{dE}{dt}\right\rangle _{N} &=&-\frac{G^{2}m(-2E\mu )^{3/2}%
}{15c^{5}\bar{L}^{7}}\left( 148E^{2}\bar{L}^{4}+732G^{2}m^{2}\mu ^{3}E\bar{L}%
^{2}+425G^{4}m^{4}\mu ^{6}\right) \ , \\
\left\langle \frac{dE}{dt}\right\rangle _{DD} &=&\frac{Gd_{1}d_{2}(-2E\mu
)^{3/2}}{15c^{5}\bar{L}^{11}}\left( C_{1}\mathcal{B}_{0}+C_{2}\mathcal{A}%
_{0}\right)
\end{eqnarray}
where the coefficients $C_{1,2}$ are 
\begin{eqnarray}
C_{1} &=&-\mu \bar{A}^{2}(948E^{2}\bar{L}^{4}+8936G^{2}E\bar{L}^{2}m^{2}\mu
^{3}+8335G^{4}m^{4}\mu ^{6})\ ,  \notag \\
C_{2} &=&708E^{3}\bar{L}^{6}+10020G^{2}E^{2}\bar{L}^{4}m^{2}\mu
^{3}+18865G^{4}E\bar{L}^{2}m^{4}\mu ^{6}+8316G^{6}m^{6}\mu ^{9}\ .
\label{coeffC}
\end{eqnarray}

\subsection{Change in the magnitude of orbital momentum}

Since there is no secular spin evolution in the 2PN order, the loss in the
magnitude $L$ under radiation reaction can be written as 
\begin{equation}
\frac{dL}{dt}\simeq \hat{\mathbf{L}}\cdot \frac{d\mathbf{J}}{dt}\ ,
\label{dLgen}
\end{equation}
where $\simeq $ denotes equality modulo spin terms, which average out due to 
$\left\langle d\mathbf{S}_{\mathbf{i}}/dt\right\rangle =0$, see \cite%
{spinspin2}. The instantaneous loss of the total angular momentum $\mathbf{J}
$ to leading order is 
\begin{equation}
\frac{d\mathbf{J}^{i}}{dt}=-\frac{2G}{5c^{5}}\epsilon
^{ijk}I^{(2)jl}I^{(3)kl}\ .
\end{equation}
Inserting $v^{2}$, $\dot{r}^{2}$ and the respective components of $\mathbf{n}
$ and $\mathbf{{d}_{i}}$ from (\ref{coords}) into (\ref{dLgen}) we obtain 
\begin{subequations}
\begin{eqnarray}
\hat{\mathbf{L}}\cdot \frac{d\mathbf{J}}{dt} &=&\left( \hat{\mathbf{L}}\cdot 
\frac{d\mathbf{J}}{dt}\right) _{N}+\left( \hat{\mathbf{L}}\cdot \frac{d%
\mathbf{J}}{dt}\right) _{DD}\ , \\
\left( \hat{\mathbf{L}}\cdot \frac{d\mathbf{J}}{dt}\right) _{N} &=&\frac{%
8G^{2}m\bar{L}}{5c^{5}\mu r^{5}}\left( 2E\mu r^{2}-3\bar{L}^{2}\right) \ , \\
\left( \hat{\mathbf{L}}\cdot \frac{d\mathbf{J}}{dt}\right) _{DD} &=&\frac{%
4Gd_{1}d_{2}}{5c^{5}\mu ^{2}\bar{L}^{3}r^{7}}\left\{ \sum_{k=1}^{3}b_{k}%
\mathcal{B}_{k}(\chi )+b_{4}\mathcal{A}_{0}\right\} \ ,  \label{instLlossDD}
\end{eqnarray}
with the coefficients $b_{k}$ given by 
\end{subequations}
\begin{eqnarray}
b_{1} &=&3\mu \bar{A}r(2GEm\mu ^{3}r^{3}-E\bar{L}^{2}\mu r^{2}-9G\bar{L}%
^{2}m\mu ^{2}r+8\bar{L}^{4})\ ,  \notag \\
b_{2} &=&3(2G^{2}Em^{2}\mu ^{5}r^{4}+22E\bar{L}^{4}\mu r^{2}-9G^{2}\bar{L}%
^{2}m^{2}\mu ^{4}r^{2}+10G\bar{L}^{4}m\mu ^{2}r-19\bar{L}^{6})\ ,  \notag \\
b_{3} &=&\mu \bar{A}r(2GEm\mu ^{3}r^{3}+3E\bar{L}^{2}\mu r^{2}-9G\bar{L}%
^{2}m\mu ^{2}r-24\bar{L}^{4})\ ,  \notag \\
b_{4} &=&\bar{L}^{4}(-18E\mu r^{2}-8Gm\mu ^{2}r+15\bar{L}^{2})\ .
\label{bbs}
\end{eqnarray}
After averaging, we obtain the secular loss in the magnitude of orbital
angular momentum: 
\begin{eqnarray}
\left\langle \frac{dL}{dt}\right\rangle &=&\left\langle \frac{dL}{dt}%
\right\rangle _{N}+\left\langle \frac{dL}{dt}\right\rangle _{DD}\ ,
\label{averLloss} \\
\left\langle \frac{dL}{dt}\right\rangle _{N} &=&-\frac{4G^{2}m(-2E\mu )^{3/2}%
}{5c^{5}\bar{L}^{4}}\left( 14E\bar{L}^{2}+15G^{2}m^{2}\mu ^{3}\right) \ , \\
\left\langle \frac{dL}{dt}\right\rangle _{DD} &=&\frac{Gd_{1}d_{2}(-2E\mu
)^{3/2}}{5c^{5}\bar{L}^{8}}\left[ D_{1}\mathcal{B}_{0}+D_{2}\mathcal{A}_{0}%
\right] \ ,
\end{eqnarray}
where the coefficients $D_{1,2}$ are given as 
\begin{eqnarray}
D_{1} &=&-6\mu \bar{A}^{2}(31E\bar{L}^{2}+90G^{2}m^{2}\mu ^{3})\ ,  \notag \\
D_{2} &=&252E^{2}\bar{L}^{4}+1200G^{2}E\bar{L}^{2}m^{2}\mu
^{3}+805G^{4}m^{4}\mu ^{6}\ .  \label{DDs}
\end{eqnarray}

\subsection{Evolution of the angles characterizing the spins under radiation
reaction}

The relative orientation of the momenta can be described by the angles $%
\kappa _{i}$ and $\gamma $, see Fig 1. Consequence of $\left\langle d\mathbf{%
S}_{\mathbf{i}}/dt\right\rangle =0,$ the angle $\gamma $ is conserved: 
\begin{equation}
\frac{d}{dt}\cos \gamma \simeq 0\ .  \label{avergammaloss}
\end{equation}%
The angles $\kappa _{i}$ were found to evolve due to both the spin-orbit 
\cite{GPV3} and the spin-spin interactions \cite{spinspin2}. Here we compute
a third contribution to their evolution, due to magnetic dipole-magnetic
dipole interaction: 
\begin{eqnarray}
\left( \frac{d}{dt}\cos \kappa _{i}\right) _{DD} &\simeq &\frac{1}{\bar{L}}(%
\mathbf{\hat{S}_{i}}-\mathbf{\hat{L}}\cos \kappa _{i})\cdot \left( \frac{d%
\mathbf{J}}{dt}\right) _{DD}  \notag \\
&=&\frac{3Gd_{1}d_{2}}{5c^{5}\mu ^{2}\bar{L}^{2}r^{7}}\sin \kappa
_{i}\sum_{j=1}^{2}\zeta _{3-j}\Bigl\{\sum_{k=1}^{3}u_{k}\left[ \sigma
_{j}\cos (k\chi +\delta _{i}+\delta _{j})-\rho _{j}\sin (k\chi +\delta
_{i}+\delta _{j})\right]   \notag \\
&&+u_{4}\sin \chi \left[ \sigma _{j}\sin (\delta _{i}-\delta _{j})-\rho
_{j}\cos (\delta _{i}-\delta _{j})\right] +u_{5}\left[ \rho _{j}\sin (\delta
_{i}-\delta _{j})+\sigma _{j}\cos (\delta _{i}-\delta _{j})\right] \Bigr\}
\label{instkappaloss}
\end{eqnarray}%
with the coefficients $u_{k}$ 
\begin{eqnarray}
u_{1} &=&-u_{3}=\mu \bar{A}r(2E\mu r^{2}-3\bar{L}^{2})\ ,  \notag \\
u_{2} &=&-2\bar{L}^{2}(Gm\mu ^{2}r+3\bar{L}^{2})\ ,  \notag \\
u_{4} &=&2\mu \bar{A}r(2E\mu r^{2}-5\bar{L}^{2})\ ,  \notag \\
u_{5} &=&2\bar{L}^{2}(4E\mu r^{2}+Gm\mu ^{2}r-5\bar{L}^{2})\ .  \label{uus}
\end{eqnarray}%
Averaging yields the following secular expression for the change of the
angle $\kappa _{i}$: 
\begin{eqnarray}
\left\langle \frac{d\kappa _{i}}{dt}\right\rangle _{DD}=\frac{%
3Gd_{1}d_{2}(-2E\mu )^{3/2}}{5c^{5}\bar{L}^{9}}\sum_{j=1}^{2}\zeta _{3-j}%
\bigl\{ &V_{1}&\left[ \sigma _{j}\cos (\delta _{i}+\delta _{j})-\rho
_{j}\sin (\delta _{i}+\delta _{j})\right]   \notag \\
&+&V_{2}\left[ \sigma _{j}\cos (\delta _{i}-\delta _{j})+\rho _{j}\sin
(\delta _{i}-\delta _{j})\right] \bigr\}\ ,  \label{averkappaloss}
\end{eqnarray}%
where $V_{1,2}$ are 
\begin{eqnarray}
V_{1} &=&5\mu \bar{A}^{2}(4E\bar{L}^{2}+7G^{2}m^{2}\mu ^{3})\ ,  \notag \\
V_{2} &=&48E^{2}\bar{L}^{4}+140G^{2}E\bar{L}^{2}m^{2}\mu
^{3}+70G^{4}m^{4}\mu ^{6}\ .  \label{VVs}
\end{eqnarray}%
Note that although the detailed expressions of the secular evolutions of $E$%
, $L$ and $\kappa _{i}$ are different from the corresponding expressions
charcterizing the gravitational quadrupole-monopole interaction \cite%
{quadrup}, the coefficients $a_{1-4},$ $b_{1-4},$ $u_{1-5}$, $C_{1,2},$ $%
D_{1,2}$ and $V_{1,2}$ are identical! This is related to the similar
structure of the respective Lagrangians.

\subsection{Evolution of the angles characterizing the magnetic moments
under radiation reaction}

The first order differential equations (\ref{averEloss}), (\ref{averLloss}),
(\ref{avergammaloss}), and (\ref{averkappaloss}) together with the algebraic
constraints (presented in detail in \cite{GPV3}): 
\begin{eqnarray}
S_{1}\sin \kappa _{1}\cos \psi _{1}+S_{2}\sin \kappa _{2}\cos \psi _{2}
&=&0\ ,  \label{ang1} \\
\cos \kappa _{1}\cos \kappa _{2}+\sin \kappa _{1}\sin \kappa _{2}\cos \left(
\psi _{2}-\psi _{1}\right) &=&\cos \gamma \ .  \label{ang2}
\end{eqnarray}
does not form a closed system, due to the presence of the angles $\lambda $, 
$\alpha _{i}$, and $\beta _{i}$ contained in $A_{0}$, $B_{0}$, and $\zeta
_{i}.$ This is a new feature to be contrasted with the spin-orbit, spin-spin
and quadrupole-monopole cases. Therefore we need the radiative evolution
equations for the above angles either.

For this purpose we remark that $\lambda $ can be expressed in term of the
other enlisted angles as 
\begin{equation}
\cos \lambda =(\rho _{2}\sigma _{1}-\rho _{1}\sigma _{2})\sin (\delta
_{1}-\delta _{2})+(\rho _{1}\rho _{2}+\sigma _{1}\sigma _{2})\cos (\delta
_{1}-\delta _{2})+\zeta _{1}\zeta _{2}\ .  \label{ang3}
\end{equation}%
A second remark is that similarly to $\left\langle d\mathbf{S}_{\mathbf{i}%
}/dt\right\rangle =0$, we expect $\left\langle d\mathbf{d}_{\mathbf{i}%
}/dt\right\rangle =0$ to hold. The assumption can be lifted by considering
any neutron star model which relates the magnetic dipole moment to other
characteristics, like the spin\footnote{%
Actual magnetar models relate the non-radiative spin precession to the
magnetic dipole-moment. Concerning the evolution of the magnetic field, they
deal only with non-radiative evolution, ranging from allowing no evolution
at all \cite{nodecay} to a non-linear magnetic field decay through the
Hall-drift \cite{Geppert}.}.

Therefore the radiative change of the angles $\alpha _{i}=\cos ^{-1}(\mathbf{%
\hat{S}_{i}\cdot \hat{d}_{i}})$ is beyond the order of accuracy of our
computation 
\begin{equation}
\frac{d}{dt}\cos \alpha _{i}\simeq 0\ ,  \label{averalphaloss}
\end{equation}%
and the only task remains to compute the radiative evolution of 
\begin{equation}  \label{cosbeta}
\cos \beta _{i}=\frac{^{\perp }\mathbf{\hat{d}}_{\mathbf{i}}\cdot \left( 
\mathbf{\hat{S}}_{\mathbf{i}}\times \mathbf{\hat{L}}\right) }{\sin \kappa
_{i}}\ ,
\end{equation}%
where by 
\begin{equation}
^{\perp }\mathbf{\hat{d}}_{\mathbf{i}}=\left( 
\begin{array}{c}
\sin\beta_i\cos\kappa_i\cos\psi_i + \cos\beta_i\sin\psi_i \\ 
\sin\beta_i\cos\kappa_i\sin\psi_i - \cos\beta_i\cos\psi_i \\ 
-\sin\beta_i\sin\kappa_i%
\end{array}%
\right)
\end{equation}%
we have denoted the unit vector aligned with the projection of the magnetic
dipole moment to the plane perpendicular to the spin of the respective
neutron star. The vector product of the projection of the magnetic dipole
moment and the direction of the spin, appearing in the subsequent
expressions, is readily obtained 
\begin{equation}
^{\perp }\mathbf{\hat{d}}_{\mathbf{i}}\times\mathbf{\hat{S}}_{\mathbf{i}} =
\left( 
\begin{array}{c}
\sin\beta_i\sin\psi_i - \cos\beta_i\cos\psi_i\cos\kappa_i \\ 
-\sin\beta_i\cos\psi_i - \cos\beta_i\sin\psi_i\cos\kappa_i \\ 
\cos\beta_i\sin\kappa_i%
\end{array}%
\right) \ .
\end{equation}%
Up to terms which do not contribute to the secular expressions 
\begin{equation}
\frac{d}{dt}\cos \beta _{i}\simeq \frac{^{\perp }\mathbf{\hat{d}}_{\mathbf{i}%
}\times\mathbf{\hat{S}}_{\mathbf{i}}} {L\left( \chi \right)\sin \kappa _{i}}%
\cdot \frac{d\mathbf{J}}{dt} -\frac{\cos \beta _{i}}{L\left( \chi \right) }%
\frac{dL}{dt}-\frac{\cos \kappa _{i}}{\sin \kappa _{i}}\cos \beta _{i}\frac{%
d\kappa _{i}}{dt}\ .  \label{betaloss}
\end{equation}%
By virtue of Eq. (\ref{dLgen}) and Eq. (\ref{cosbeta}) we find that to the
leading order in the $DD$-terms, Eq. (\ref{betaloss}) becomes: 
\begin{equation}
\frac{d}{dt}\left(\sin \kappa _{i}\cos \beta _{i}\right) \simeq \frac{1}{%
\bar{L}}\left\{ ^{\perp }\mathbf{\hat{d}}_{\mathbf{i}}\times \mathbf{\hat{S}}%
_{\mathbf{i}} - \left[ \left(^{\perp }\mathbf{\hat{d}}_{\mathbf{i}}\times 
\mathbf{\hat{S}}_{\mathbf{i}}\right) \cdot \mathbf{\hat{L}}\right] \mathbf{%
\hat{L}} \right\} \left( \frac{d\mathbf{J}}{dt}\right)_{DD} \ .
\label{betaloss2}
\end{equation}%
As $(d\mathbf{J}/dt)_N\sim\mathbf{\hat{L}}$ (for which the expression in
brackets acts as a projector) we find that similarly to the angles $\kappa
_{i}$, the radiative evolution of the angles $\beta _{i}$ receives no
Newtonian contribution. Employing Eq. (\ref{instkappaloss}), the detailed
expression of Eq.(\ref{betaloss2}) turns out to be 
\begin{eqnarray}
\frac{d}{dt}\cos \beta _{i} &=&- \frac{3Gd_{1}d_{2}}{5c^{5}\mu ^{2}\bar{L}%
^{2}r^{7}} \frac{\sin\beta_i}{\sin\kappa_{i}} \sum_{j=1}^{2}\zeta
_{3-j}\Bigl\{\sum_{k=1}^{3}u_{k}\left[ \rho _{j}\cos (k\chi +\delta
_{i}+\delta_{j}) +\sigma _{j}\sin(k\chi +\delta _{i}+\delta _{j}) \right] 
\notag \\
&&-u_{4}\sin \chi \left[ \sigma_{j}\cos (\delta _{i}-\delta _{j}) +\rho
_{j}\sin (\delta _{i}-\delta _{j})\right] +u_{5}\left[ \sigma _{j}\sin
(\delta_{i}-\delta _{j}) -\rho _{j}\cos (\delta _{i}-\delta _{j})\right]
\Bigr\}  \label{instbetaloss}
\end{eqnarray}
and its average over a radial period gives 
\begin{eqnarray}
\left\langle \frac{d\beta _{i}}{dt}\right\rangle _{DD}=-\frac{%
3Gd_{1}d_{2}(-2E\mu )^{3/2}}{5c^{5}\bar{L}^{9}\sin\kappa_i}
\sum_{j=1}^{2}\zeta _{3-j}\bigl\{ &V_{1}&\left[ \rho_{j}\cos (\delta
_{i}+\delta _{j}) +\sigma _{j}\sin (\delta _{i}+\delta _{j})\right]  \notag
\\
&+&V_{2}\left[ \sigma _{j}\sin(\delta _{i}-\delta _{j}) -\rho _{j}\cos
(\delta _{i}-\delta _{j}) \right] \bigr\}\ .  \label{averbetaloss}
\end{eqnarray}
Note that the coefficients appearing in the instantaneous and averaged loss
of the angles $\beta_i$ agree with the coefficients (\ref{uus}) and (\ref%
{VVs}) in the radiative change of $\kappa_i$.

\section{Concluding remarks}

We have given a closed system of differential equations governing the
evolution of a set of dynamical and geometrical variables under the magnetic
dipole - magnetic dipole contribution to the radiation reaction. These are
the energy, magnitude of orbital angular momentum, angles between the spins
and orbital angular momentum and angles between the dipole moments and the
respective spins. These evolutions add to previously derived PN, 2PN, SO, SS
and QM contributions. For sufficiently strong magnetic fields this new
contribution is of second PN order.

As a check of the above results we compare the circular orbit limit of the
energy loss with the previous result of Ioka and Taniguchi \cite{IT}.
Imposing the circularity condition in an average sense (as described in Ref. 
\cite{quadrup}), the following relations hold 
\begin{equation}
\bar{E}_{N}=E-\bar{E}_{DD}=-\frac{Gm\mu }{2r_{0}}\ ,\quad \bar{L}_{N}^{2}=%
\bar{L}^{2}=Gm\mu ^{2}r_{0}\ .  \label{circEL}
\end{equation}
Here $r_{0}$ is the radius of the unperturbed circular orbit and the angular
average $\bar{E}_{DD}$ of (\ref{Edd}) is 
\begin{equation}
\bar{E}_{DD}=\frac{d_{1}d_{2}\mathcal{A}_{0}}{2r_{0}^{3}}\ .
\end{equation}
Inserting $E$ and $\bar{L}$ from Eqs. (\ref{circEL}) into the expression of
the energy loss (\ref{averEloss}) we obtain the radiative loss of energy for
the previously defined circular orbit: 
\begin{equation}
\left\langle \frac{dE}{dt}\right\rangle =-\frac{32G^{4}m^{3}\mu ^{2}}{%
5c^{5}r_{0}^{5}}\left( 1-\frac{12d_{1}d_{2}\mathcal{A}_{0}}{Gm\mu r_{0}^{2}}%
\right) \ .
\end{equation}
Comparison with the corresponding result of \cite{IT} is achieved by
remarking that Eq. (\ref{A0}) can be written in the form 
\begin{equation}
\mathcal{A}_{0}=3(\mathbf{\hat{L}\cdot \hat{d}_{1}})(\mathbf{\hat{L}\cdot 
\hat{d}_{2}})-\mathbf{\hat{d}_{1}\cdot \hat{d}_{2}}\ ,
\end{equation}
and by computing the averaged radius of the quasi-circular orbit in terms of
the radius of the Newtonian circular orbit $r_{0}$: 
\begin{equation}
\left\langle r\right\rangle =r_{0}\left( 1+\frac{3d_{1}d_{2}\mathcal{A}_{0}}{%
2m\mu r_{0}^{2}}\right) \ .
\end{equation}
After the required series expansion we find that our energy loss, when
specialized to circular orbits and expressed in terms of $\left\langle
r\right\rangle $, is in perfect agreement with the energy loss contribution
due to magnetic dipole - magnetic dipole interaction, given in Eq.(16) of
Ref. \cite{IT}.

\section{Acknowledgments}

We thank Krist\'of Petrovay and Peter Biermann for useful references. This
research was supported by OTKA no. T031724 and T034615 grants.


\begin{thebibliography}{99}
\bibitem{LIGO} A. Abramovici et al., Science \ \textbf{256}, 325 (1992).

\bibitem{VIRGO} C. Bradaschia et al., Nucl. Instrum. Methods{\ Phys. Res. A}
\ \textbf{289}, 518 (1990).

\bibitem{GEO} J. Hough, in \textit{Proceedings of the Sixth Marcell
Grossmann Meeting}, edited by H. Sato and T. Nakamura (World Scientific,
Singapore, 1992), p. 192.

\bibitem{TAMA} K. Kuroda et al., in \textit{Proceedings of International
Conference on Gravitational Waves: Sources and Detectors}, edited by I.
Ciufolini and F. Fidecaro (World Scientific, Singapore, 1997), p. 100.

\bibitem{LISA} P. Bender, I. Ciufolini, K. Danzmann, W. M. Folkner, J.
Hough, D. Robertson, A. R\"{u}diger, M. C. W. Sandford, R. Schilling, B. F.
Schutz, R. Stebbins, T. Sumner, P. Touboul, S. Vitale, H. Ward, and W.
Winkler: \textit{LISA: Pre-Phase A Report (MPQ 208)} (Max-Planck Institut f%
\"{u}r Quanten\-optik, Garching, Germany, 1996).

\bibitem{LISA1} B. F. Schutz, in \textit{Proceedings of the 1997 Alpbach
Summer School on Fundamental Physics in Space}\emph{, }edited by A. Wilson%
\emph{, } gr-qc/9710079.

\bibitem{Blanchet} L. Blanchet, \textit{Post-Newtonian Theory and its
Application}, to appear in the Proc. of the 12th Workshop on General
Relativity and Gravitation, ed. M. Shibata, gr-qc/0304014 (2003).

\bibitem{IBBH} P.R. Brady, J.D.E. Creighton, and K.S. Thorne \prd \textbf{58}%
, 061501 (1998).

\bibitem{Buonanno1} A. Buonanno, Y. Chen, and M. Vallisneri, \prd \textbf{67}%
, 024016 (2003).

\bibitem{Peters} P. C. Peters, Phys. Rev. \textbf{136}, B1224 (1964).

\bibitem{QuSha} G. D. Quinlan and S. L. Shapiro, \apj \textbf{321}, 199
(1987).

\bibitem{HiBe} D. Hills and P. L. Bender, Astrophys. J. Lett. \textbf{445},
L7 (1995).

\bibitem{PoMa} K. Martel and E. Poisson, \prd \textbf{60}, 124008 (1999).

K. Martel, \textit{Detection of Gravitational Waves from Eccentric Compact
Binaries}, gr-qc/9908043.

\bibitem{Biermann} H. Falcke, P. L. Biermann, W. J. Duschl and P. G. Mezger,
Astron. Astrophys. \textbf{270}, 102 (1993).

\bibitem{nospin1} L.~Bildsten and C.~Cutler, Astrophys. J. \textbf{400}, 175
(1992).

\bibitem{nospin2} C.S.~Kochanek, Astrophys. J. \textbf{398}, 234 (1992).

\bibitem{magnetars} C. Kouveliotou, {\ et al.}, Nature \textbf{393}, 235
(1998).

\bibitem{PetersMathews} P. C. Peters and J. Mathews, Phys. Rev. \textbf{131}%
, 435 (1963).

\bibitem{Buonanno2} A.~Buonanno, Y.~Chen and M.~Vallisneri,  \prd \textbf{67}%
, 104025 (2003).

\bibitem{Grandclement} P.~Grandcl\'{e}ment and V.~Kalogera, \prd \textbf{67}%
, 082002 (2003).

\bibitem{GI} A. Gopakumar and B. R. Iyer, \prd \textbf{56}, 7708 (1997).

\bibitem{BOC} B. M. Barker and R. F. O'Connell, \prd \textbf{12}, 329 (1975).

\bibitem{ACST} T. A. Apostolatos, C. Cutler, G. J. Sussman and K. S. Thorne, %
\prd\ \textbf{49}, 6274 (1994).

\bibitem{KWW} L. Kidder, C. Will, and A. Wiseman, \prd \textbf{47}, 4183
(1993).

\bibitem{Kidder} L. Kidder, \prd \textbf{52}, 821 (1995).

\bibitem{RS} R. Rieth and G. Sch\"{a}fer, Class. Quantum Grav.\ \textbf{14},
2357 (1997).

\bibitem{GPV3} L. \'{A}. Gergely, Z. Perj\'{e}s, and M. Vas\'{u}th, \prd\ 
\textbf{58}, 124001 (1998).

\bibitem{spinspin1} L. \'{A}. Gergely, \prd \textbf{61}, 024035 (2000).

\bibitem{spinspin2} L. \'{A}. Gergely, \prd \textbf{62}, 024007 (2000).

\bibitem{quadrup} L. \'{A}. Gergely and Z. Keresztes, \prd\ \textbf{67},
024020 (2003).

\bibitem{Poisson} E. Poisson, \prd \textbf{57}, 5287 (1998).

\bibitem{IT} K. Ioka and T. Taniguchi, Astrophys. J. \textbf{537}, 327
(2000).

\bibitem{param} L. \'{A}. Gergely, Z. Perj\'{e}s, and M. Vas\'{u}th,
Astrophys. J. Suppl. Ser. \textbf{126}, 79\textbf{\ }(2000).

\bibitem{nodecay} T. Regimbau and J. de Freitas Pacheco, Astron. and
Astrophys. \textbf{374}, 182 (2001).

\bibitem{Geppert} U. Geppert and M. Rheinhardt, astro-ph/0207065, to appear
in Astron. and Astrophys.
\end{thebibliography}
\end{document}